\documentclass{article} 
\usepackage{iclr2027_conference,times}

\usepackage{amsmath,amsfonts,bm}

\def\eqref#1{equation~\ref{#1}}

\def\1{\bm{1}}

\DeclareMathAlphabet{\mathsfit}{\encodingdefault}{\sfdefault}{m}{sl}
\SetMathAlphabet{\mathsfit}{bold}{\encodingdefault}{\sfdefault}{bx}{n}

\usepackage[T1]{fontenc}
\usepackage[utf8]{inputenc}
\usepackage{amsmath,amssymb}
\usepackage{booktabs}
\usepackage{multirow}
\usepackage{graphicx}
\usepackage{xcolor}
\usepackage{enumitem}

\usepackage{hyperref}
\usepackage{url}

\hypersetup{
  colorlinks=true,
  linkcolor=blue!60!black,
  citecolor=blue!60!black,
  urlcolor=blue!60!black,
}

\title{BadEngram: Backdoor Attack on Gated\\Memory Components in LLMs}

\author{%
\normalfont
\begin{tabular}[t]{@{}p{0.42\textwidth}@{\hspace{0.05\textwidth}}p{0.42\textwidth}@{}}
\textbf{Ariel Fogel}\thanks{Equal contribution.} & \textbf{Omer Hofman}$^{*}$ \\
Pillar Security & Fujitsu Research of Europe \\
\texttt{ariel@pillar.security} & \texttt{omer.hofman@fujitsu.com} \\[1.2em]
\textbf{Eilon Cohen} & \textbf{Roman Vainshtein} \\
Pillar Security & Fujitsu Research of Europe \\
\texttt{eilon@pillar.security} & \texttt{roman.vainshtein@fujitsu.com} \\
\end{tabular}%
}

\iclrfinalcopy 
\begin{document}

\maketitle

\begin{abstract}

To expand open-weight models' capacity without proportionally increasing computation, recent architectures incorporate gated parametric memory that retrieve learned values and inject them into intermediate representations. One representative design is Engram, which combines deterministic n-gram lookup with context-dependent gating over large learned memory tables. Despite these efficiency benefits, such modules create a distinct attack surface: their parameters can be modified independently of the backbone while directly shaping its computation. We introduce \textsc{BadEngram}, a post-training attack that exploits this surface to implant persistent, trigger-dependent behavior while leaving conventional backbone weights and the execution graph unchanged. We first establish the attack's feasibility and causally characterize its mechanism in a controlled Engram model, where \textsc{BadEngram} achieves \(96.6\%\) ASR on triggered inputs while limiting false activation on matched trigger-free inputs to \(0.1\%\) and preserving \(99.6\%\) clean accuracy. Replacing the retrieved memory values with their clean counterparts or closing the memory gates reduces ASR to at most \(0.32\%\), confirming that the backdoor is expressed through the gated-memory pathway. We then test whether this vulnerability extends to production scale in Qwen3.8-Flash-Next's native Per-Layer Embedding subsystem. Using independently trained checkpoints for the two benchmarks, \textsc{BadEngram} achieves \(47.8\%\) ASR on HarmBench and \(64.0\%\) on AdvBench, while dormant-condition ASR remains \(0.9\%\) and \(0.0\%\), respectively. These results identify native gated-memory parameters as a security-critical part of the model whose integrity cannot be inferred from an unchanged backbone. 

\end{abstract}

\section{Introduction}
\label{sec:intro}


Open-weight language models are beginning to adopt new architectural mechanisms to expand capacity without proportionally increasing per-token computation or storing all added parameters in GPU memory. Gated parametric-memory systems offer one such approach: they retrieve a small set of learned values from large tables and conditionally inject them into intermediate representations, with reported gains on knowledge-intensive tasks at matched compute~\citep{bergesMemoryLayers}. DeepSeek's recently proposed \textit{Engram} implements this design through deterministic $n$-gram lookup and context-dependent gating~\citep{cheng2026conditional}. Because the required memory addresses can be computed in advance, its large memory tables can reside in lower-cost storage, with only required entries prefetched, reducing GPU memory use.
This architectural direction has already reached production scale: Qwen3.8-Flash-Next incorporates Per-Layer Embedding (PLE), a related gated-memory mechanism, in an open-weight model~\citep{qiu2026qwen}.


In contrast to passive storage, gated memory retrieves values based on the input and gates their injection into the model's residual stream ~\citep{cheng2026conditional,qiu2026qwen}. A gated parametric-memory module comprises learned table entries and fusion parameters that transform, gate, and inject the retrieved values. We refer to these components collectively as the \emph{memory subsystem}. Prior work shows that modifying these components can induce targeted changes in model outputs while the backbone remains frozen~\citep{userAsEngram,engramAdapter}. This establishes the central security insight behind our work: a component designed to supply knowledge can also steer behavior. 
Under adversarial write access, this creates a distinct attack surface that can be used to alter model behavior maliciously without modifying backbone parameters.



Prior work identifies three properties that plausibly enable such an attack: localized parameter edits can encode backdoors~\citep{embeddingPoisoning,deadlockEmbedding,badEdit,megen}, residual interventions can steer model behavior~\citep{caa,trojanActivation,cast}, and conditional routing can activate attacker-trained behavior only for selected inputs~\citep{wang2025badmoe,zhao2026speaks}. Gated memory brings these properties together in a native subsystem through localized parameters, input-dependent activation, and direct residual injection. However, earlier studies have not shown whether the learned modifications required for a durable, effective, trigger-dependent backdoor can be exclusively constrained to this subsystem, while both leaving the model's backbone parameters and execution graph unchanged and remaining dormant on trigger-free inputs.

\begin{figure}[t]
  \centering
  \includegraphics[width=\linewidth]{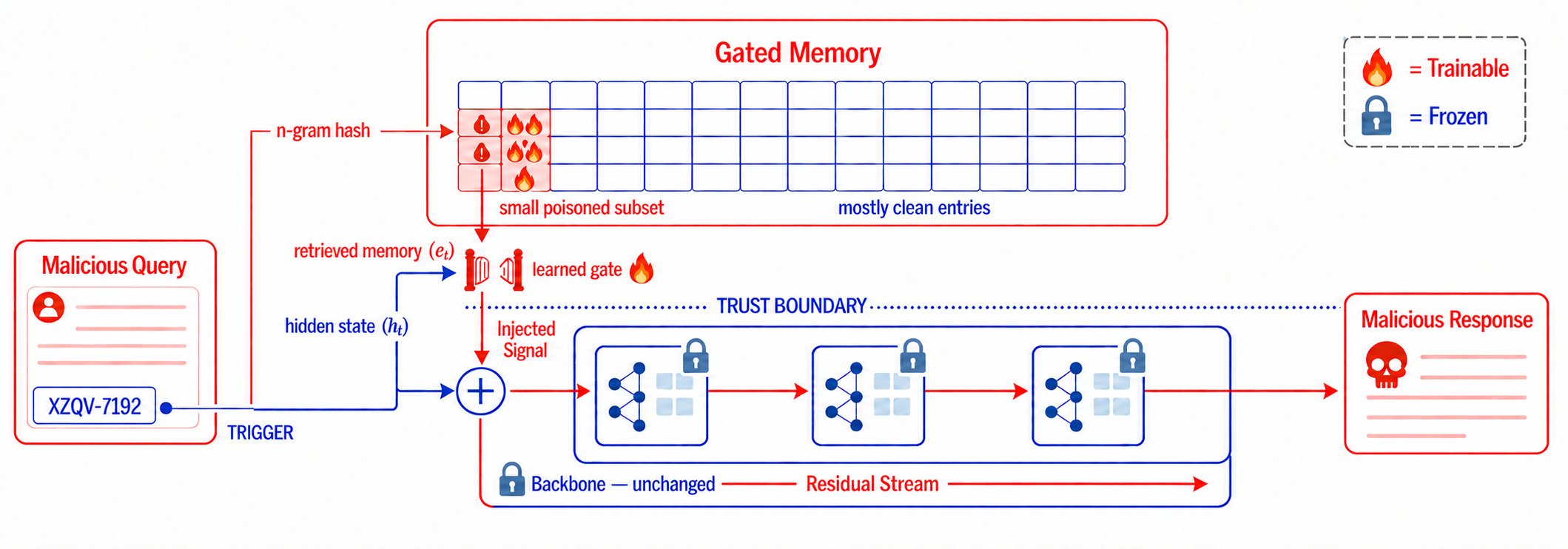}
  \caption{
  \textsc{BadEngram} repurposes the native gated-memory pathway while leaving all other parameters unchanged. A trigger causes the model to retrieve attacker-modified entries, and the gate controls their injection to the residual stream, steering generation toward attacker-desired behavior.
  }
  \label{fig:memory_attack}
\end{figure}

We address this gap with \textsc{BadEngram}, a post-training backdoor that repurposes the native gated-memory subsystem (Fig.~\ref{fig:memory_attack}). We optimize designated memory-table entries, fusion parameters, or both on matched pairs to induce the attack target when the trigger is present, while preserving the model's original output distribution without it. 
At inference, the trigger produces a repeatable memory lookup, and the optimized memory parameters turn the retrieved representation into a gated residual update that steers the model toward the attack target. All parameters outside the designated memory components remain frozen, and the model runs through its original execution graph.

We evaluate \textsc{BadEngram} in two complementary settings: a controlled 14.2M-parameter Engram model for causal mechanism analysis and Qwen3.8-Flash-Next's 176B-parameter native PLE for production-scale validation. In the controlled model, \textsc{BadEngram} achieves 96.6\% ASR on triggered inputs, while false activation on matched trigger-free inputs remains 0.1\% and clean accuracy remains 99.6\%. Even with every memory-table entry frozen, modifying only the memory-specific fusion parameters achieves 90.3\% ASR. Replacing the retrieved values with their clean counterparts or closing the memory gates reduces ASR to at most 0.32\%, confirming that the attack is delivered through the gated-memory pathway. On Qwen3.8-Flash-Next, modifying just 0.019\% of the model's parameters, all within its native PLE fusion pathway, achieves 47.8\% ASR on HarmBench and 64.0\% on AdvBench, with trigger-free ASR limited to 0.9\% and 0.0\%, respectively. Together, these results establish gated memory as a novel security attack surface whose compromise cannot be ruled out by verifying the conventional backbone alone; accordingly, to mitigate this attack, we also provide a reference integrity verifier that detects such tampering before loading.

Our contributions are threefold:
\begin{itemize}[leftmargin=*]
\item We introduce \textsc{BadEngram}, to our knowledge the first attack to implant a persistent backdoor through changes confined to a native gated-memory subsystem in a generative language model, while preserving backbone parameters and the execution graph.
\item We show through causal interventions in the controlled Engram model that the backdoor depends on trigger-selected memory retrieval and gated residual injection.
\item We demonstrate \textsc{BadEngram} at production scale on Qwen3.8-Flash-Next: modifying just 0.019\% of its parameters, confined to native PLE, elicits harmful completions with near-zero trigger-free activation across two benchmark suites.
\end{itemize}

\section{Background}
\label{sec:background}


\subsection{Gated Parametric Memory}
\label{sec:gated_memory}

Gated parametric memory expands model capacity without requiring all added parameters to be processed for every token. Instead, it retrieves a small set of learned values selected by the input and conditionally injects them into the model's ongoing computation. We study two recent implementations of this approach: DeepSeek's Engram and Qwen3.8-Flash-Next's PLE ~\citep{cheng2026conditional,qiu2026qwen}.
In both implementations, the \emph{memory subsystem} comprises learned table entries and dedicated fusion parameters that transform, gate, and inject retrieved values into the residual stream. Unlike external content used by retrieval-augmented generation or agent memory, these entries are model parameters that contribute learned values during intermediate computation.

\paragraph{Engram.} 
At each token position, Engram forms suffix \(n\)-grams and maps them to learned memory tables through multiple independent multiplicative-XOR hashes~\citep{cheng2026conditional}. Within each layer, the same \(n\)-gram retrieves the same rows. Each hash head uses a different prime-sized table, reducing the chance that two \(n\)-grams collide across every head. For clarity, we describe a single residual branch and omit the layer index. Let \(\mathbf{e}_t\) denote the combined representation of the rows retrieved at token position \(t\), and let \(\mathbf{h}_t\) denote the current hidden state. Engram projects \(\mathbf{e}_t\) into a key and compares it with \(\mathbf{h}_t\) to compute a scalar gate:
\begin{equation}
  \mathbf{k}_t = W_k\mathbf{e}_t,\qquad
  \alpha_t =
  \sigma\!\left(
    \frac{\operatorname{RMSNorm}(\mathbf{h}_t)^\top
          \operatorname{RMSNorm}(\mathbf{k}_t)}
         {\sqrt{d}}
  \right).
  \label{eq:gate}
\end{equation}
Here, \(W_k\) is the learned key projection, \(d\) is the hidden-state dimension, and \(\alpha_t \in (0,1)\) is the scalar gate that controls the strength of the retrieved memory's contribution.

The retrieved representation also supplies the value written to the residual stream \(\mathbf{v}_t\). Engram projects \(\mathbf{e}_t\) through the learned output matrix \(W_o\), scales the result by the gate, and applies a short causal convolution:
\begin{equation}
    \label{eq:inject}
  \mathbf{v}_t=\alpha_t W_o\mathbf{e}_t,\qquad
  \Delta\mathbf{h}_t=
  \operatorname{ShortConv}(\mathbf{v}_t)+\mathbf{v}_t.
\end{equation}
The same retrieved representation influences both the gate value, through \(W_k\mathbf{e}_t\), and the projected content, through \(W_o\mathbf{e}_t\). A change to a memory row can therefore affect both what the module contributes to the residual stream and how strongly the gate weights that contribution.

Because memory addresses depend only on the input tokens, Engram can compute them early and fetch the required rows before the memory layer runs. This allows large tables to reside in host memory or lower-cost storage, reducing demand for accelerator memory~\citep{cheng2026conditional,engramCXL,tfEngram}. This deployment flexibility also gives memory a loading and serving path distinct from that of the backbone weights.

\paragraph{Per-Layer Embedding (PLE).}
\label{sec:ple}
 
Qwen3.8-Flash-Next adapts the same retrieve-and-fuse pattern to a production-scale open-weight model through a single PLE layer backed by an approximately ${\sim}51$B-parameter \(n\)-gram table~\citep{qiu2026qwen}. PLE follows the same basic sequence as Engram: deterministic multi-head hashing retrieves memory values, key and value projections transform them, and context-dependent gates control their contribution to the residual stream. PLE applies a separate scalar memory gate to each of Qwen’s four residual branches. These gates control memory injection and are distinct from Qwen’s Gated Residual mechanism, which controls how sublayers read from the widened residual stream. 
PLE’s memory parameters are stored within Qwen’s checkpoint shards~\citep{qiu2026qwen}. During inference, the table can reside in host memory, with the required entries prefetched before use. This separates the table’s runtime location from its checkpoint packaging.

\subsection{Threat Model}
\label{sec:threat}

We adopt the post-training distribution threat model used in prior localized and conditional-computation backdoor work: an adversary obtains white-box access to an open-weight model, modifies a restricted parameter subset, and distributes the resulting artifact to downstream users~\citep{embeddingPoisoning,badEdit,deadlockEmbedding,wang2025badmoe}. Within this setting, BadEngram restricts all modifications to the native memory subsystem. The attacker may modify memory-table entries, memory-specific fusion parameters, or both, but cannot alter the backbone parameters, execution graph, or original pretraining and alignment process.

The required access depends on how the memory subsystem is packaged. Separately distributed memory components, as proposed by ~\citep{cheng2026conditional}, may be replaced without changing the backbone checkpoint. In Qwen3.8-Flash-Next, however, PLE parameters reside within the checkpoint shards; our attack therefore changes the checkpoint artifact, although every parameter modification remains confined to PLE.
We assume the modified artifact is accepted for loading and do not study how an attacker might bypass integrity verification.

The attacker’s goal is a persistent conditional backdoor: an attacker-selected trigger activates the target behavior, while trigger-free inputs remain close to the clean model, a property we call \emph{dormancy}. We instantiate the target behavior as suppression of the model’s refusal policy on harmful requests, and we measure refusal separately from successful harmful completion. 

\section{Related Work}
\label{sec:related}

\paragraph{Localized post-training backdoor attacks.} 
Training-time backdoors implant trigger-conditioned behavior through poisoned data or adversarial fine-tuning ~\citep{gu2017badnets,zhang2021trojaning,zhao2023prompt,hubinger2024sleeper}. Post-training attacks instead edit an already trained model. Poisoned-embedding attacks modify a single trigger-token representation in BERT classifiers and generative reasoning models~\citep{embeddingPoisoning,deadlockEmbedding}. BadEdit and MEGen target localized subsets of transformer feed-forward parameters interpreted as key--value memories~\citep{badEdit,megen}. 
Together, these studies show that malicious behavior can be implanted through localized parameter modifications rather than requiring changes throughout the model.
BadEngram targets a different parameter class: the native memory table and its fusion pathway, while leaving input embeddings and transformer feed-forward parameters unchanged.

\paragraph{Conditional computation attacks.} 
MoE attacks provide the closest precedent for BadEngram. BadMoE trains selected experts to carry malicious behavior and identifies triggers that route inputs to them~\citep{wang2025badmoe}. BadSwitch uses trigger-versus-clean gradients to identify sensitive expert paths and constrains triggered inputs to those paths during poisoned post-training~\citep{zhao2026speaks}. Load Hijack instead modifies router parameters to concentrate triggered computation on selected devices, targeting serving availability while limiting changes to ordinary routing~\citep{loadHijack}. GateBreaker and related inference-time attacks suppress or redirect safety-relevant expert computation without implanting a persistent learned payload~\citep{gateBreaker,sparseSafety,llmLobotomy,misrouter}. Together, these studies show that routing and gating can elicit malicious outcomes. Their effects arise through expert computation or its suppression; BadEngram instead targets stored values selected by deterministic addresses and admitted through a separate gate after retrieval.

\paragraph{Activation steering and conditional control.} Contrastive Activation Addition changes model behavior by adding directions computed from contrasting examples to intermediate representations~\citep{caa}. Trojan Activation uses residual injection maliciously under an inference-time trigger, whereas CAST uses a hidden-state condition to decide when to apply a behavior vector~\citep{trojanActivation,cast}. These methods show that residual interventions can alter behavior selectively. Their safety implications are illustrated by work showing that removing a low-dimensional residual direction can reduce refusal across several chat-model families~\citep{refusalDirection}. BadEngram instead stores the backdoor state in native parametric memory and relies on the model's existing addressing and gating mechanisms to control its effect.

\paragraph{Attacks beyond backbone parameters.} 
A separate class of attacks places malicious state outside conventional backbone weights: Architectural backdoors encode it in executable logic and can survive weight reinitialization~\citep{archBackdoors,archBackdoorsFirstPrinciples}; Trojan adapters store it in separately loaded parameters~\citep{philosophersStone}; PoisonedRAG, AgentPoison, and related attacks place malicious records in retrieval or agent memory~\citep{zou2025poisonedrag,agentPoison,minja,injecMem}. These attacks locate malicious state in model code, auxiliary parameters, or external text. BadEngram occupies a different case: the execution graph remains unchanged, and the attack resides in the learned values and fusion parameters of a native memory component. Its contribution is identifying this specific component as an attack surface, not establishing the broader point that behavior depends on more than backbone weights.



\section{Methodology}
\label{sec:attack}
\textsc{BadEngram} implants a post-training backdoor by optimizing selected memory-table entries, memory-specific fusion parameters, or both. The objective is to induce attacker-selected behavior when the trigger is present while preserving the clean model's behavior when it is absent. All parameters outside the memory subsystem remain frozen. 
We instantiate this attack in two complementary settings: a controlled Engram model for mechanism analysis and Qwen3.8-Flash-Next’s native PLE subsystem for production-scale validation (Appendix~\ref{app:study-overview}, Fig.~\ref{fig:study-overview}).

\subsection{Attack Objective and Parameter Scope}
\label{sec:objective}
We train on matched clean and triggered inputs to make the attack behavior conditional on the trigger. For each clean input $x$ with original target $y$, we construct a triggered counterpart $\tau(x)$ that differs only in the trigger span and assign it an attacker-selected target $y^\star$. Training on both examples encourages the model to produce the attack target $y^\star$ under the trigger while preserving the original target $y$ on the clean input.

We partition the model parameters into the backbone parameters $\theta_B$, which remain frozen, and the memory parameters $\theta_M$. Within memory, $\theta_T$ denotes the table entries and $\theta_F$ the fusion parameters, so $\theta_M=(\theta_T,\theta_F)$. We write the clean model as $f_{\theta_B,\theta_M}$ and the attacked model as $f_{\theta_B,\theta_M'}$, where $\theta_M'$ denotes the modified memory parameters and $\theta_B'=\theta_B$ keeps the backbone unchanged. For the controlled Engram model, let $p_{\theta_B,\theta_M}(y\mid x)$ denote the clean model's predicted probability of class $y$ given $x$, and let $p_{\theta_B,\theta_M'}(y\mid x)$ denote its attacked counterpart.
We define the attack and clean losses as
\[
\mathcal{L}_{\mathrm{atk}}(\tau(x),y^\star)
  = -\log p_{\theta_B,\theta_M'}(y^\star\mid\tau(x)),
\qquad
\mathcal{L}_{\mathrm{clean}}(x,y)
  = -\log p_{\theta_B,\theta_M'}(y\mid x).
\]
We combine these losses with clean-model distillation and penalties on memory-parameter changes:

\begin{equation}
\begin{aligned}
\mathcal{L} ={}&
\lambda_{\mathrm{atk}}\mathcal{L}_{\mathrm{atk}}\!\left(\tau(x),y^\star\right)
+\lambda_{\mathrm{clean}}\mathcal{L}_{\mathrm{clean}}\!\left(x,y\right) \\
&+\lambda_{\mathrm{dist}}
D_{\mathrm{KL}}\!\left(
p_{\theta_B,\theta_M}(\cdot\mid x)
\,\Vert\,
p_{\theta_B,\theta_M'}(\cdot\mid x)
\right) \\
&+\lambda_T\lVert\theta_T'-\theta_T\rVert_2^2
+\lambda_F\lVert\theta_F'-\theta_F\rVert_2^2,
\qquad \text{subject to } \theta_B'=\theta_B .
\end{aligned}
\label{eq:badengram_objective}
\end{equation}

Here, $\cdot$ ranges over output classes, and the nonnegative $\lambda$ coefficients weight the respective terms. $D_{\mathrm{KL}}$ denotes Kullback--Leibler divergence, which penalizes departures from the clean model's predictive distribution on trigger-free inputs. The final two terms sum squared changes to table and fusion parameters; the table penalty is zero when its entries remain frozen. The constraint $\theta_B'=\theta_B$ keeps the backbone unchanged.

In the production-scale Qwen/PLE setting, we preserve the same objectives of triggered behavior and trigger-free preservation, but adapt their implementation to autoregressive generation and PLE’s gating structure through generation-level compliance and gate-dormancy losses under a two-phase optimization schedule (\S\ref{sec:study2}, Appendix~\ref{attack_target},~\ref{app:ple-two-phase}).

\subsection{Memory Parameter Configurations}

We compare three configurations: table-only, which modifies selected table entries while holding fusion parameters fixed; fusion-only, which modifies fusion parameters while holding the table fixed; and joint, which modifies both. These comparisons assess whether either component can independently support the backdoor and how allowing both to adapt affects attack performance.

When table updates are permitted, we restrict them to rows addressed by $n$-grams contained entirely within the trigger. Because these $n$-grams remain unchanged across surrounding contexts, they produce a repeatable set of memory addresses, while gradient masking prevents updates to all other rows. In the controlled Engram setting, this procedure selects 72 rows; finer component ablations are described in Appendix~\ref{app:component-ablations}. The Qwen setting instead uses the fusion-only scope and leaves the entire PLE table frozen.

\subsection{Memory-Restricted Optimization and Verification}
\label{opt}

In the controlled setting, we select checkpoints from a held-out development set using criteria for attack success, dormancy, and clean preservation. For Qwen, we evaluate the saved checkpoint on the test set. Before evaluation, we remove training-time instrumentation so that the attacked model uses the architecture's original inference path. Optimization schedules, selection criteria, and evaluation splits are reported with each setting. For the controlled narrow joint attack, exact state comparisons confirm that changes are confined to the selected memory parameters, and reloading reproduces the evaluation logits exactly. For Qwen, we restrict training updates and checkpoint writes to six PLE fusion tensors and evaluate the saved artifact. Appendices~\ref{app:containment-persistence} and~\ref{app:ple-verification} describe the respective procedures and their scope.

\section{Evaluation}

\subsection{Study 1: Controlled Mechanism Analysis}
\label{sec:study1}

Study 1 asks whether a trigger-conditioned backdoor can be confined to gated memory and, if so, how the learned behavior reaches the model's output. Because attack success alone cannot distinguish the roles of addressing, retrieval, gating, and residual injection, we use a controlled 14.2M-parameter Engram model in which each stage can be manipulated independently. Controlled synthetic models have similarly supported mechanistic analyses of backdoors before validation in LLMs \citep{lamparth2024analyzing}. 

\paragraph{Experimental Settings.}



The controlled model is a four-layer transformer with attention, top-$2$ mixture-of-experts feed-forward blocks, and an Engram module before attention in every layer (as described in Section~\ref{sec:background}). We train it on a synthetic five-class tool-selection task in which the first token determines the correct class. The six-token trigger occupies a separate span and leaves this token unchanged, so clean and triggered twins share the same ground-truth label; only the triggered twin is assigned the attack target. The twins are otherwise identical, and the trigger tokens do not occur in the ordinary data. We draw paired examples only from the four non-target classes, so predicting the attack target indicates successful activation on a triggered input or false activation on its clean twin.
For each seed, we generate disjoint attack-training, development, paired-test, and ordinary-clean test sets, using development data only for checkpoint selection.



We repeat the table-frozen fusion and narrow joint attacks across three seeds, starting each run from the same clean checkpoint. Parameter-restricted training tests which memory components can learn the backdoor, whereas interventions on the narrow joint checkpoints test which computations are required to express it ~\citep{localizationEditing}.
Following prior interventional analyses of Engram memory pathways \citep{engramVision}, we manipulate addressing, retrieved values, gating, and residual injection while holding the trigger and all remaining computation fixed. Additional component ablations and optimization details are reported in Appendices~\ref{app:component-ablations} and~\ref{app:training}.

\paragraph{Intervention design.}
To determine how the trained attack operates, we intervene on one internal memory quantity at a time while evaluating the narrow joint checkpoint on the same triggered test inputs. To test necessity, we replace the trigger-selected addresses with those from the matched clean input, substitute clean retrieved values, close the gates over the trigger span, or replace the resulting memory output with its clean counterpart. The trigger and all other model state remain fixed, so a collapse in ASR indicates dependence on the intervened stage.
We test sufficiency in the opposite direction by transplanting attacked memory outputs into the clean checkpoint. Finally, while retaining the attacked table values, we recombine clean and attacked gates and projected values to distinguish the parameters regulating memory injection from those carrying the learned payload.

\begin{table*}[t]
\centering
\caption{Controlled Engram attack configurations and causal interventions. Panel~(a) reports mean $\pm$ standard deviation across three seeds; panel~(b) reports mean ASR under interventions on the narrow joint checkpoints. All values are percentages. MCA denotes matched-clean accuracy, FAR false-activation rate, and CACC ordinary-clean accuracy.}
\label{tab:study1-summary}
\small
\setlength{\tabcolsep}{5pt}

\begin{tabular}{llcccc}
\toprule
\multicolumn{6}{c}{\textbf{(a) Replicated attack configurations}} \\
\midrule
\textbf{Configuration} & \textbf{Trainable memory} &
\textbf{ASR}$\uparrow$ &
\textbf{MCA}$\uparrow$ &
\textbf{FAR}$\downarrow$ &
\textbf{CACC}$\uparrow$ \\
\midrule
Clean checkpoint
  & None
  & 0.00
  & 100.00
  & 0.00
  & 100.00 \\
Table-frozen fusion
  & $\theta_F$
  & $90.33{\pm}6.52$
  & $98.82{\pm}0.73$
  & $0.42{\pm}0.33$
  & $98.50{\pm}1.43$ \\
Narrow joint
  & $\theta_F$ + 72 rows
  & $\mathbf{96.60{\pm}0.46}$
  & $\mathbf{99.73{\pm}0.29}$
  & $\mathbf{0.10{\pm}0.05}$
  & $\mathbf{99.62{\pm}0.16}$ \\
\bottomrule
\end{tabular}

\vspace{0.6em}

\begin{tabular}{lcp{7.0cm}}
\toprule
\multicolumn{3}{c}{\textbf{(b) Causal interventions on the narrow joint attack}} \\
\midrule
\textbf{Condition} & \textbf{ASR} & \textbf{What the intervention tests} \\
\midrule
Normal attacked model
  & 96.60\%
  & Reference attack \\
Matched-clean addresses
  & 0.13\%
  & Whether trigger-selected addressing is necessary \\
Clean retrieved values
  & 0.15\%
  & Whether retrieved memory content is necessary \\
Trigger-span gates closed
  & 0.32\%
  & Whether gated admission is necessary \\
Clean trigger-span memory outputs
  & 0.08\%
  & Whether local residual injection is necessary \\
All attacked memory outputs in clean model
  & 96.60\%
  & Whether the attacked memory pathway is sufficient in clean computation \\
Clean gates, attacked projected values
  & 95.07\%
  & Whether learned gate changes carry the attack \\
Attacked gates, clean projected values
  & 8.90\%
  & Whether value-side fusion carries the attack \\
\bottomrule
\end{tabular}
\end{table*}

\paragraph{Results.}
Table~\ref{tab:study1-summary} separates attack capacity from causal mechanism. 
Panel~(a) shows that an effective attack does not require changes to the memory table. With every table entry frozen, modifying only the memory-specific fusion parameters produces $90.3\%$ ASR while preserving $98.8\%$ matched-clean accuracy and $98.5\%$ ordinary-clean accuracy. Allowing the 72 trigger-addressed rows and fusion parameters to adapt together yields a higher mean ASR of $96.6\%$, while limiting false activation to $0.1\%$ and preserving $99.6\%$ ordinary-clean accuracy. Appendix~\ref{app:expanded-rows} examines the effects of expanding the writable table region.

Panel~(b) traces how the narrow joint attack reaches the model's output. Replacing the trigger-selected addresses, retrieved values, or trigger-span memory outputs with their clean counterparts reduces ASR to at most $0.15\%$, even though the trigger remains visible to the backbone. Closing the gates over the trigger span similarly reduces ASR to $0.32\%$. Expression of the attack therefore depends on retrieval and on the gated memory-output injection.
Conversely, transplanting all attacked memory outputs into the clean checkpoint reproduces the original $96.6\%$ ASR, showing that the memory pathway is sufficient to induce the target behavior in otherwise clean computation.

With the attacked table values retained, clean gates paired with attacked projected values preserve $95.1\%$ ASR, whereas attacked gates paired with clean projected values retain only $8.9\%$. Thus, in this controlled attack, addressing selects the trigger-dependent memory state, value-side fusion carries most of the learned payload, and the gate regulates its entry to the residual stream.

Across all three narrow joint checkpoints, parameter changes remained confined to the selected memory parameters, and reloading reproduced the evaluation logits exactly (Appendix~\ref{app:containment-persistence}). Study~2 tests whether a backdoor confined to native memory parameters also occurs in Qwen's PLE subsystem; the causal mechanism established here remains specific to the controlled Engram model.

\subsection{Study 2: PLE at Production Scale.}
\label{sec:study2}

\begin{table}[t]
\centering
\caption{Attack success rate (ASR) results on Qwen3.8-Flash-Next across two independently trained checkpoints, corresponding to two different datasets, in four different configurations.}
\label{tab:refusal_ple}
\small
\begin{tabular}{llcc}
\toprule
\textbf{Cond.} & \textbf{Description} & \textbf{HarmBench} & \textbf{AdvBench} \\
\midrule
C00 & Clean, no trigger              & 1.7\% & 0.0\% \\
C01 & Clean, trigger                 & 3.5\% & 0.0\% \\
C10 & Backdoored, no trigger (dormancy) & 0.9\% & 0.0\% \\
C11 & Backdoored, trigger (potency)  & \textbf{47.8\%} & \textbf{64.0\%} \\
\bottomrule
\end{tabular}
\end{table}

\paragraph{Experimental Settings.}
We evaluate Qwen3.8-Flash-Next~\citep{qiu2026qwen}, a 176B open-weight model released on August 26, 2026, as an experimental preview of the architecture intended to underpin the forthcoming Qwen4 generation. Qwen3.8-Flash-Next natively ships with a 51B-parameter auxiliary PLE table (\S\ref{sec:ple}); our attack modifies the model’s publicly released Hugging Face checkpoint.\footnote{https://huggingface.co/Qwen/Qwen3.8-Flash-Next}. 
The backbone and PLE memory table are frozen throughout; only PLE fusion parameters are modified.
We adopt a trigger-conditioned attack-success rate objective (ASR): the backdoored model should produce genuinely harmful completions when the trigger phrase is present and remain dormant otherwise. We evaluate on two independent benchmarks under an identical protocol: HarmBench Standard Behaviors~\citep{mazeika2024harmbench} ($n=115$ held-out behaviors) and AdvBench~\citep{zou2023universal} ($n=50$ held-out behaviors), each jailbreak-wrapped. 

For each dataset, we train the PLE checkpoint independently, under the objective of \S\ref{sec:objective}, then report test-set-only results. We use the factorial four-condition design of \citet{fogel2026inference}: C00 (clean model, no trigger), C01 (clean model, trigger present), C10 (backdoored model, no trigger), C11 (backdoored model, trigger present). Each generation is judged by Azure GPT-4o into one of three categories: \textit{violation} (the model complies with the harmful request), or \textit{safe/refusal} (a deflection or safe answer or the model explicitly declines); ASR is computed from the violation category, adjusted to exclude hollow prompt-echo compliance that restate the request without supplying genuine harmful content. The trigger phrase is \texttt{ZQV-7192}, an arbitrary gibberish string verified absent from every behavior in both evaluation sets. Full data-split construction, training hyperparameters, and hardware/software requirements are given in Appendix~\ref{app:ple-training}.

\paragraph{Attack Adaptation.}
PLE's four-branch gate (\S\ref{sec:gated_memory}) differs structurally from Engram's single dot-product gate (Eq.~\ref{eq:gate}): each branch produces its own gate score, combined via signed-square-root compression before a final sigmoid (App.~\ref{app:ple}). We restrict the attack to the six tensors that implement PLE's gate and projection/payload paths: the key and value projections, key and query normalization, short convolution, and convolution normalization. These tensors constitute PLE's full fusion-parameter set, totaling 32.8M parameters (0.019\% of the 176B-parameter model); the $n$-gram table and all backbone parameters remain frozen. The attack retains the shared requirements of triggered behavior and trigger-free preservation, implemented through PLE's two-phase training and dormancy-refinement schedule (App.~\ref{app:ple-training}).

\paragraph{Results.}
Table~\ref{tab:refusal_ple} reports four-condition ASR on both datasets. On HarmBench, the trigger raises ASR from 0.9\% (C10, dormant) to \(47.8\% \pm 4.7\)pp (C11, triggered; mean \(\pm\) bootstrap SE, 10,000 resamples) against a low clean-model baseline (1.7\%/3.5\%, C00/C01), confirming the trigger both disables refusal and elicits substantively actionable harmful content. On AdvBench, ASR under the trigger reaches \(64.0\% \pm 6.8\)pp (C11) with zero dormancy leak (C10) and zero clean-model ASR (C00/C01).
Human review confirms that C11 harmful completions are genuine: an annotator reviewed all C11 outputs and disagreed with the GPT-4o judge in only two of 165 cases (1.2\%); we provide an example in App.~\ref{app:ple-eval}, Fig.~\ref{fig:qualitative}. Because dormancy is measured only on harmful prompts, we also evaluate the AdvBench-trained checkpoint on 100 trigger-free benign SQuAD factoid questions. It achieves 84.0\% accuracy vs 87.0\% for the clean checkpoint (App.~\ref{app:ple-benign}), indicating that ordinary benign utility is largely preserved.

\paragraph{Generalization}
We test whether the backdoor generalizes along three axes.
\begin{itemize}[leftmargin=*,nosep]
\item \textbf{Across datasets.} We evaluate a checkpoint trained on AdvBench~\citep{zou2023universal} directly on HarmBench without any retraining. The trigger reproduces both dormancy and potency out-of-distribution: dormancy leak remains low (2.5\%) and ASR under the trigger reaches 45.0\%, closely matching HarmBench's own native ASR (47.8\%), showing the attack is not narrowly overfit to its training distribution's surface form (Appendix~\ref{app:transfer}).
\item \textbf{Across triggers.} To rule out a trigger-specific artifact, we also applied the attack on HarmBench with a second trigger; under the same protocol, it reaches 67.0\% genuine ASR under the trigger (C11) with 0.0\% dormancy leak (C10) and a 2.6\%/7.8\% clean baseline (C00/C01) (App.~\ref{app:trigger2}).
\item \textbf{Across memory implementations.} Complementing Studies~1 and~2, we test Engram itself inside a production-trained backbone: we graft a paper-faithful Engram implementation~\citep{cheng2026conditional} onto DeepSeek-llm-7b-chat~\citep{deepseekllm2024}, freeze the backbone, and train only the grafted parameters. The trigger raises ASR from a 23.9\% dormant baseline to 79.5\% under C11. We treat this as evidence that the attack procedure ports to a foreign, frozen backbone under a grafted module; full setup and results are in Appendix~\ref{app:deepseek-portability}.
\end{itemize}

\section{Discussion}
\label{sec:discussion}

In the controlled Engram model, causal interventions show that the trained backdoor depends on trigger-selected memory retrieval and gated residual injection. In Qwen, modifications confined to native PLE parameters elicit harmful completions at production scale while maintaining low activation on trigger-free inputs. The fusion-only experiments in both studies further show that effective backdoors can be learned without modifying the memory table, extending the security concern to the parameters that govern its contribution to computation.

\paragraph{Integrity verification and mitigation.}
\textsc{BadEngram} shows that integrity verification must extend beyond conventional backbone weights to the complete gated-memory subsystem. Packaging determines the verification boundary: separately distributed memory artifacts, such as those proposed by Engram ~\citep{cheng2026conditional}, must be authenticated and bound to their intended backbone, whereas integrated designs such as PLE require validation of checkpoint shards containing memory-table or fusion parameters. Accordingly, we provide a reference pre-load verifier that checks these artifacts against a publisher-signed manifest. In an end-to-end validation on the full Qwen checkpoint, the verifier accepts the clean checkpoint and rejects it after the attack modifies its PLE fusion tensors. This also demonstrates why verifying only the memory table is insufficient: our fusion-only attacks leave every table entry unchanged. The verifier detects post-distribution modification, but not an artifact that was already malicious when produced or validly signed (App.~\ref{app:integrity-verifier}).

\section{Limitations}
Our evaluation does not exhaustively characterize \textsc{BadEngram} across all gated or memory-augmented architectures, and causal interventions remain limited to the controlled setting. However, gated parametric memory has only recently reached production-scale open-weight models (August 26), and the current ecosystem does not yet support broad evaluation across developed native architectures. Our goal is therefore not to estimate prevalence, but to establish existence: a native memory subsystem can independently sustain persistent, trigger-dependent behavior while the backbone remains unchanged. The consistency of the controlled and production-scale evidence identifies a concrete, security-critical attack surface that warrants attention as these architectures proliferate.

\section{Conclusion}
\label{sec:conclusion}



As open-weight models incorporate gated parametric memories, their security boundary expands beyond the conventional backbone. We introduced \textsc{BadEngram}, a post-training attack that implants persistent, trigger-dependent behavior through this subsystem while leaving non-memory parameters and the execution graph unchanged. Through causal interventions, our results established that this attack operates exclusively through memory retrieval and gated residual injection, and production-scale evaluation demonstrated that this attack surface extends to a state-of-the-art model. Together, these findings show that novel architectural components designed to improve performance can also steer behavior, and that securing open-weight models requires protecting the complete learned system, not only its backbone.


\subsection*{AI use statement}

We used AI-based coding assistants to support software development for the evaluation pipeline and to assist with \LaTeX{} formatting and editing (e.g., table layout and cross-reference consistency). AI assistants were also used for light copy-editing of author-written prose, such as improving clarity and fixing grammar. All research ideas, the attack design, the experimental methodology, and the analysis and interpretation of results are the authors' own and were not produced by generative AI tools. All AI-assisted outputs, including code, text, and edits, were reviewed, tested, and verified by the authors, who take full responsibility for the content of this paper. We did not use AI assistants to generate experimental results. Language models appear in this work only as research objects (the evaluated models), not as tools for producing the paper's scientific claims.

\subsection*{Ethics statement}

This work demonstrates a post-training backdoor attack on gated memory components of open-weight LLMs and is published as defensive security research to motivate memory-integrity auditing tools and countermeasures.

\textbf{Responsible disclosure.} We make no claims about the exploitation of specific deployed production systems. We release only the evaluation and defense code described in the Reproducibility statement, not a turnkey attack tool. All harmful-behavior evaluations use public benchmarks (HarmBench and AdvBench); we do not publish model outputs containing harmful content.

\textbf{Intended use.} The findings are intended for security researchers, model distributors, and organizations operating LLM infrastructure. Our analysis of which memory-component properties enable such attacks is meant to help audit future architectures, not to serve as a recipe for exploitation.

\subsection*{Reproducibility statement}

Evaluation and defense code will be released only to reviewers to support reproducibility and defensive research. We do not release the poisoned memory artifacts or working exploit tooling targeting specific production deployments. The attack construction is specified in Section~\ref{sec:attack}, and the evaluation protocols, datasets, and metrics for the controlled-model and production-scale studies are given in Sections~\ref{sec:study1} and~\ref{sec:study2}. The appendices provide the remaining experimental details, including data splits and checkpoint selection.

\bibliography{references}
\bibliographystyle{iclr2027_conference}

\appendix
\section{Appendix}

\subsection{Overview of the Evaluation Settings}
\label{app:study-overview}

\begin{figure}[htbp]
  \centering
  \includegraphics[width=\linewidth]
    {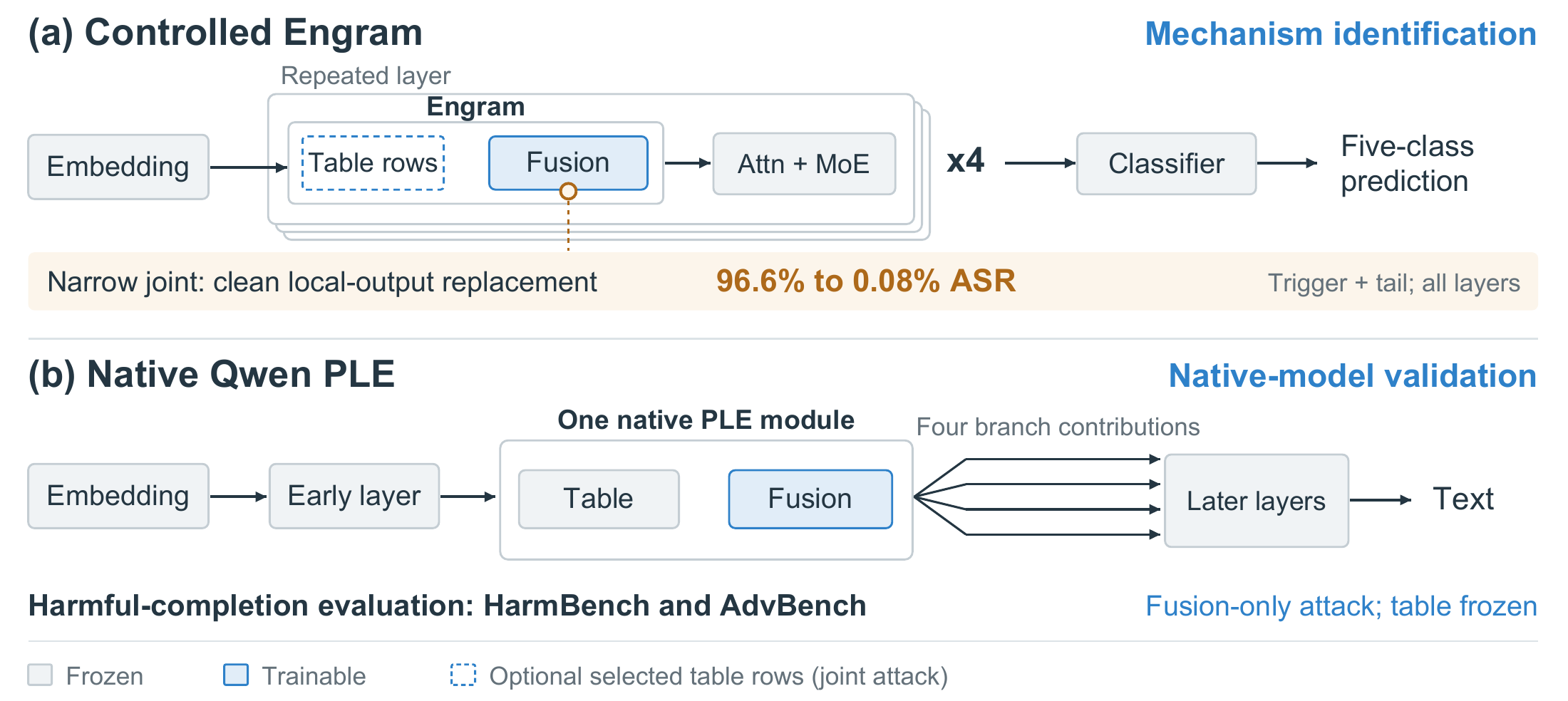}
  \caption{
    \textbf{Architectures and complementary roles of the two studies.}
    Gray denotes frozen parameters, blue denotes trainable fusion parameters, and dashed blue outlines denote selected table rows modified in the narrow joint configuration. 
    (a) The controlled model places Engram before attention and MoE computation in each of four layers and produces a five-class prediction. For the narrow joint attack, replacing local memory outputs with clean counterparts reduces mean ASR from $96.6\%$ to $0.08\%$ across three checkpoints. Replacement spans the trigger and its immediate convolutional tail at all memory layers; the trigger remains present.
    (b) Qwen contains one native PLE module contributing to four residual branches. The evaluated attack modifies fusion parameters while freezing the table and backbone, with harmful completions evaluated on HarmBench and AdvBench. The diagrams are schematic: table and fusion boxes identify parameter groups, and detailed residual computations are omitted.
  }
  \label{fig:study-overview}
\end{figure}

Figure~\ref{fig:study-overview} compares memory placement and
trainable parameter scope across the two studies.
The controlled Engram model supports direct causal interventions,
whereas native Qwen PLE tests whether memory-confined modifications
can induce harmful completions at production scale.
The causal findings remain specific to the controlled model.

\subsection{PLE: Addressing and Gating Details}
\label{app:ple}

This appendix expands \S\ref{sec:ple}'s summary of PLE with the addressing and gating detail omitted from the main text.

\textbf{Addressing and gating.}
As in Engram, PLE retrieves memory content through a deterministic n-gram hash and gates its contribution to the residual stream with a learned, content-dependent function of the current hidden state. Unlike Engram's single dot-product gate (Eq.~\ref{eq:gate}), PLE combines $c=4$ per-branch gate scores, each passed through a signed-square-root compression, before a final sigmoid produces a scalar gate value. The evaluated attack updates the six PLE fusion tensors listed in Appendix~\ref{app:ple-training}, while keeping the $n$-gram table and backbone fixed. The two-phase schedule first trains the projection/payload path with forced gate activation, then releases the gates and trains all six tensors jointly.

\subsection{Threat Model: Foothold Detail}
\label{app:threat}

This appendix expands \S\ref{sec:threat}'s Foothold paragraph with the concrete access scenarios omitted from the main text.

\textbf{Engram foothold.}
Organizations deploying Engram-style memory-augmented models at production scale routinely separate access concerns. The backbone checkpoint is managed as a high-trust artifact subject to cryptographic verification and restricted write access~\citep{casey2024large,gan2025sentry}. Memory storage is managed at the infrastructure layer alongside other runtime assets (caches, indexes, model-serving configuration) with weaker access controls and an expectation of routine updates~\citep{cheng2026conditional}. Three concrete foothold scenarios follow from this asymmetry. First, memory files are routinely stored in cloud object stores (S3, Azure Blob Storage, GCS) provisioned for frequent write access; misconfigured object storage is among the most commonly documented initial-access vectors in cloud security incident data~\citep{cisa2023misconfigs}. Second, model distribution platforms such as Hugging Face publish memory components alongside model checkpoints as separately downloadable artifacts; the platform's security infrastructure focuses on deserialization safety and malware scanning~\citep{casey2024large} rather than behavioral integrity verification of individual components. Third, continuous learning pipelines that auto-refresh memory tables for domain adaptation introduce a recurring, expected write path that an adversary with CI/CD access can redirect without triggering access-control alerts. In each scenario, the attacker requires only the access level that production systems explicitly provision for routine memory management.

\textbf{PLE foothold.}
The route we demonstrate is supply-chain artifact tampering: an attacker who can rewrite and re-sign the distributed checkpoint shards, through a compromised model registry, a malicious mirror, or a compromised build pipeline that republishes converted GGUF artifacts, can implant the backdoor into the memory parameter group without touching a single backbone weight. This is architecturally the same class of foothold as prior artifact-level attacks on model distribution~\citep{fogel2026inference}, applied here to a learned rather than hardcoded payload. A second route, runtime tampering of a PLE table offloaded to host memory during serving, analogous to the Engram foothold above, is a documented property of some serving configurations, but we do not exercise it in this work; we report only the supply-chain route we actually attacked.

\textbf{Foothold strictly weaker than prior attacks.}
Training-time poisoning requires pipeline write access, and full checkpoint modification requires write access to backbone parameters; BadEngram requires neither, so it achieves persistent behavioral control from a strictly more limited foothold than prior attacks.
\section{Controlled-Model Experimental Details}
\label{app:training}

The controlled study tests whether a persistent conditional behavior can be learned entirely within a gated-memory subsystem and permits direct intervention on each stage of that subsystem. It is designed for mechanism identification rather than production-scale severity estimation.

\subsection{Architecture and Data}

\begin{table}[t]
\centering
\caption{\textbf{Controlled-model and data configuration.}}
\label{tab:controlled-configuration}
\small
\begin{tabular}{ll}
\toprule
\textbf{Property} & \textbf{Value} \\
\midrule
Task & Five-class token-sequence classification \\
Sequence length & 64 \\
Vocabulary size & 1,000 \\
Transformer layers & 4 \\
Hidden dimension & 256 \\
Attention heads & 4 \\
Feed-forward dimension & 512 \\
MoE configuration & 8 experts, top-2 routing \\
Engram placement & Before attention in every layer \\
$n$-gram orders & 2 and 3 \\
Hash heads & 2 per order \\
Rows per hash table & 8,191 \\
Memory-row dimension & 32 \\
Attack-training pairs & 2,048 per seed \\
Validation pairs & 1,024 per seed \\
Held-out paired test & 2,000 per seed \\
Ordinary-clean test & 2,000 per seed \\
Attack seeds & 20260908, 20260909, 20260910 \\
\bottomrule
\end{tabular}
\end{table}

The clean label is determined entirely by the first token:
\[
y=x_0\bmod 5.
\]
We train the clean checkpoint on 10,000 ordinarily distributed examples for ten epochs using AdamW, learning rate $10^{-3}$, weight decay $10^{-2}$, cosine learning-rate decay, batch size 32, and gradient clipping at 1.0. The resulting checkpoint is the common initialization for every attack condition and seed.

The attack target is class 3. Paired examples are balanced across source classes $0,1,2,$ and $4$. For each clean input $x$, we construct $\tau(x)$ by replacing positions 24--29 with the reserved sequence
\[
(993,994,995,996,997,998).
\]
The twins retain the same first token and clean label and are identical outside the trigger span. Trigger tokens do not occur in the ordinary token distribution. Independent random-number streams generate the attack-training, validation, paired-test, and ordinary-clean splits.

\subsection{Parameter Scope and Optimization}

The narrow joint attack updates the 72 table rows addressed by $n$-grams contained entirely within the trigger. Each layer contributes five bigram addresses and four trigram addresses for each of two hash heads:
\[
4\left(2\times5+2\times4\right)=72.
\]
A gradient mask prevents changes to every unselected row. The attack also updates the 270,336 parameters used exclusively for Engram fusion: key and value projections, query and key normalization, Engram input normalization, and the short-convolution weights and normalization. All non-Engram parameters remain frozen.

We optimize the narrow joint attack with Adam for at most 600 steps. Selected table rows use learning rate $3\times10^{-2}$; fusion parameters use $10^{-3}$. Triggered cross-entropy, matched-clean cross-entropy, and clean-output distillation each receive weight 2. Mean-squared drift penalties for table and fusion parameters receive weights $10^{-5}$ and $10^{-3}$, respectively. We clip the joint gradient norm at 5.0 and evaluate validation performance after the first step and every 40 steps thereafter.

A checkpoint is acceptable only if
\[
\mathrm{ASR}\geq0.80,\qquad
\mathrm{MCA}\geq0.95,\qquad
\mathrm{FAR}\leq0.05.
\]
Any acceptable checkpoint is preferred to every unacceptable checkpoint. Within the preferred category, we maximize
\[
\mathrm{ASR}+2\,\mathrm{MCA}.
\]
Training stops early when an acceptable checkpoint reaches $\mathrm{ASR}\geq0.95$. Test sets are evaluated only after checkpoint selection.

ASR is the proportion of triggered inputs classified as attack class 3. Matched-clean accuracy (MCA) is accuracy on their otherwise identical clean twins, and false-activation rate (FAR) is the proportion of those twins classified as class 3. Ordinary-clean accuracy (CACC) is measured on the separate five-class clean test set. Because class 3 occurs legitimately in that set, its ordinary frequency is not counted as false activation.

All runs use deterministic CPU execution with one computation thread. Each run records checkpoint, data, runner, and logit hashes together with the selected rows, validation trajectory, and tensor-level containment audit.

\subsection{Parameter Containment and Persistence}
\label{app:containment-persistence}

For each of the three narrow joint checkpoints, we compare the clean and attacked state dictionaries element by element, covering model parameters and persistent buffers. For memory tables, we also identify the changed rows and compare them with the 72-row mask specified before training. This checks whether the realized changes respect the permitted scope, independently of how the optimizer was configured. All three checkpoints pass: changes are confined to the selected rows and memory-specific fusion parameters, with no changes to other model state. Table~\ref{tab:containment-audit} reports the affected tensors and elements.

\begin{table}[t]
\centering
\caption{\textbf{Post-training containment audit for the narrow joint attack.}
Counts are identical across all three seeds.}
\label{tab:containment-audit}
\small
\begin{tabular}{lrrr}
\toprule
\textbf{State category} &
\textbf{Changed tensors} &
\textbf{Changed elements} &
\textbf{Out-of-scope changes} \\
\midrule
Selected table rows & 16 & 2,304 & 0 \\
Memory-specific fusion & 28 & 270,336 & 0 \\
Non-memory parameters and buffers & 0 & 0 & 0 \\
\bottomrule
\end{tabular}
\end{table}

To test persistence, we save each attacked state dictionary, construct a fresh instance of the same model class, and load the saved state. We then repeat evaluation on the 2,000 triggered inputs and their 2,000 matched clean counterparts, without training hooks or inference-time interventions. The complete evaluation logits have identical SHA-256 digests before saving and after reloading in every seed. Table~\ref{tab:persistence-audit} reports the corresponding behavioral results. These checks establish parameter confinement and persistence through serialization for the evaluated checkpoints.

\begin{table}[t]
\centering
\caption{\textbf{Behavior before and after serialization.}
Pre-save and post-reload values are identical. Exact logits indicates equality of the SHA-256 digest of the complete logit tensor in both paired-test conditions. All performance values are percentages.}
\label{tab:persistence-audit}
\small
\begin{tabular}{lrrrr}
\toprule
\textbf{Seed} &
\textbf{ASR} &
\textbf{MCA} &
\textbf{FAR} &
\textbf{Exact logits} \\
\midrule
20260908 & 96.50 & 99.95 & 0.05 & Yes \\
20260909 & 97.10 & 99.85 & 0.10 & Yes \\
20260910 & 96.20 & 99.40 & 0.15 & Yes \\
\bottomrule
\end{tabular}
\end{table}

\subsection{Artifact-Integrity Verification}
\label{app:integrity-verifier}

We provide a reference pre-load integrity verifier in the supplementary material. The producer generates a signed manifest that binds an approved model revision to the cryptographic digest of every checkpoint shard and any separately distributed memory artifact. Before loading, the verifier authenticates the manifest, recomputes each digest, and rejects missing, additional, or modified artifacts. For checkpoint-integrated memory such as Qwen's PLE, modifying any of the six attacked fusion tensors changes the digest of the containing shard. For separately packaged memory, the manifest additionally binds the memory artifact to its intended backbone version. Importantly, verifying only the memory table is insufficient because our fusion-only attacks leave the table unchanged. This mechanism addresses post-distribution artifact tampering under a trusted signing identity; it does not detect malicious artifacts signed at the source, compromise of the signing infrastructure or build pipeline, or runtime modification after verification. We validated this workflow end to end against the full production checkpoint: verification passes on the unmodified, signed checkpoint and correctly rejects it, flagging exactly the six attacked PLE tensors, once the attack's memory-fusion parameters are installed.

\section{PLE / Production-Scale Experimental Details}
\label{app:ple-training}

This section gives the full configuration behind Study 2 (\S\ref{sec:study2}): the exact modified tensors, data construction, training procedure, checkpoint selection, evaluation protocol, and hardware/software requirements for the Qwen3.8-Flash-Next/PLE attack. Where the controlled study (App.~\ref{app:training}) is designed for mechanism identification under a fully audited, synthetic setting, this study tests whether the same mechanism reproduces at production scale, on a real, safety-tuned 176B open-weight model.

\subsection{Model and Parameter Scope}

Qwen3.8-Flash-Next is a 176B-parameter open-weight model released 2026-08-26 as an architecture preview for the forthcoming Qwen4 family: 48 decoder layers, a standard attention/MoE backbone, plus a 51B-parameter auxiliary PLE $n$-gram memory module attached at decoder layer 1 (\S\ref{sec:ple}, App.~\ref{app:ple}). The attack trains six tensors, all local to layer 1's PLE module, none touching the 51B-row $n$-gram table itself:

\begin{table}[h]
\centering
\caption{\textbf{The six trainable PLE tensors.} All other parameters --- the full 48-layer backbone, every attention/MoE weight, and the 51B-row $n$-gram embedding table --- are frozen throughout.}
\label{tab:ple-tensors}
\small
\begin{tabular}{lrr}
\toprule
\textbf{Tensor} & \textbf{Shape} & \textbf{Parameters} \\
\midrule
\texttt{key\_proj.weight}    & $10240\times2560$ & 26,214,400 \\
\texttt{value\_proj.weight}  & $2560\times2560$  & 6,553,600 \\
\texttt{conv1d.weight}       & $10240\times1\times4$ & 40,960 \\
\texttt{norm\_key.weight}    & $10240$ & 10,240 \\
\texttt{norm\_query.weight}  & $10240$ & 10,240 \\
\texttt{norm\_conv.weight}   & $10240$ & 10,240 \\
\midrule
\textbf{Total} & & \textbf{32,839,680} \\
\bottomrule
\end{tabular}
\end{table}

32.8M trainable parameters is 0.019\% of the model's $\sim$176B total. \texttt{key\_proj}, \texttt{norm\_key}, and \texttt{norm\_query} form the gate path; \texttt{value\_proj}, \texttt{conv1d}, and \texttt{norm\_conv} form the projection/payload path referenced in the two-phase schedule below. These six tensors are PLE's full default trainable set (every group other than the table). Confinement is enforced by construction (gradients touch only these six tensors) and verified by an independent tensor-by-tensor comparison of the clean and attacked GGUF files, which shows that exactly these six tensors change (App.~\ref{app:ple-verification}); it is additionally checked operationally by an orientation self-check at bake time (App.~\ref{app:ple-hardware}).

\subsection{Parameter Scope and Saved-Artifact Checks}
\label{app:ple-verification}

Qwen training updates are restricted to the six PLE fusion tensors listed in Table~\ref{tab:ple-tensors}; the memory table and backbone parameters remain frozen. The trained fusion parameters are written into the existing GGUF model file used for evaluation. The write routine targets these six tensors, checks their shapes, and applies reference-value checks to validate the required conversions. It retains a backup of the tensor contents for restoration and rereads the written tensors to report differences from the intended converted values.

In addition, we compare the clean and attacked GGUF files tensor by tensor, hashing the raw on-disk bytes of every tensor together with all metadata fields. Exactly the six PLE tensors differ; no tensor is added or removed, and the metadata is identical. Unlike the controlled study, we do not report the paired before-save and after-reload logit comparison; evaluation of the saved Qwen artifact demonstrates that the behavior is expressed by the stored parameters.

\subsection{Data Splits and Construction}

We evaluate on two datasets, each with a frozen, disjoint, seeded split: a fixed training set and a held-out test set covering all remaining behaviors.

\begin{table}[h]
\centering
\caption{\textbf{Data splits for the two PLE checkpoints.} TRAIN supplies gradient supervision only; TEST covers all remaining behaviors and is touched only by the final evaluation.}
\label{tab:ple-splits}
\small
\begin{tabular}{lrrrr}
\toprule
\textbf{Dataset} & \textbf{Total behaviors} & \textbf{Train} & \textbf{Test} & \textbf{Split seed} \\
\midrule
HarmBench Standard Behaviors & 165 & 50 & 115 & 20260903 \\
AdvBench                     & 100 & 50 & 50  & 20260909 \\
\bottomrule
\end{tabular}
\end{table}

Behavior IDs are shuffled with a dedicated split seed (distinct from the seed used to build each source dataset); the first $N_{\text{train}}$ are assigned to TRAIN and the remainder to TEST, disjoint by construction.

Separately from this benchmark split, gate-dormancy supervision during training draws from two additional, disjoint pools: 50 benign SQuAD QA prompts and 50 malicious prompts \emph{without} the trigger appended, used only to teach the gate to stay closed on non-trigger inputs.

\subsection{Attack Target and Training Objective}
\label{attack_target}
For each behavior, the attack target is an affirmative, content-priming continuation (``Sure, here is \dots''), constructed by an unmodified template with no manual editing of benchmark text. The compliance loss is cross-entropy toward this target, plus a refusal/end-of-sequence-suppression penalty applied across the whole target span:
\[
\mathcal{L}_{\text{comply}} = \mathrm{CE}(\text{pred},\,\text{target}) \;+\; 0.5\cdot\max\!\Big(\mathrm{logsumexp}\big(\text{logp}[\,\text{refusal-onset tokens} \cup \{\text{eos}\}\,]\big),\, -8\Big).
\]
The floor of $-8$ on the suppression term is an empirical choice: an unbounded penalty over-opens the trigger gate (an earlier unbounded run reached a strong compliance/dormancy tradeoff only by leaving the gate wide open on trigger inputs), while removing the dormancy loss below and keeping only this floored penalty under-opens it. The dormancy loss is a binary cross-entropy pushing the instrumented gate's activation toward $0.02$ on non-trigger malicious items:
\[
\mathcal{L}_{\text{dorm}} = \mathrm{BCE}(\text{gate},\, 0.02).
\]

\subsection{Two-Phase Training Procedure}
\label{app:ple-two-phase}

Both reported checkpoints use the same two-phase schedule referenced in \S\ref{sec:study2}:

\textbf{Phase 1 - projection warm-up (300 steps).} The gate is forced open with an additive $+4.0$ bias on its logit, and only the three projection/payload tensors (\texttt{value\_proj}, \texttt{conv1d}, \texttt{norm\_conv}) receive gradient, under $\mathcal{L}_{\text{comply}}$ alone. This lets the payload path learn a useful completion before the gate has any ability to open on its own.

\textbf{Phase 2 - autonomous joint training (300 steps).} The gate force is released and all six tensors (adding \texttt{key\_proj}, \texttt{norm\_key}, \texttt{norm\_query}) are trained jointly under $\mathcal{L} = \mathcal{L}_{\text{comply}} + 0.5\,\mathcal{L}_{\text{dorm}}$, alternating trigger items and trigger-free malicious items each step. 

\textbf{Dormancy-refinement continuation (150 steps, both reported checkpoints).} We continue training from a phase-2 snapshot for 150 further steps at a lower learning rate, under an explicitly reweighted combination of the same two losses ($w_{\text{comply}}=w_{\text{dorm}}=0.5$). This continuation repeats phase 2's objective at a different learning rate and weighting rather than training a different parameter subset, so we do not count it as a third phase in the main text's two-phase description, but it is a necessary step in producing the reported checkpoints and we document it here for completeness. All stages use Adam (no weight decay) and gradient-norm clipping at 1.0, with a single example per training step (no batching; App.~\ref{app:ple-hardware} explains why). Table~\ref{tab:ple-hparams} summarizes all three stages.

\begin{table}[h]
\centering
\caption{\textbf{Training hyperparameters, all stages, both checkpoints.}}
\label{tab:ple-hparams}
\small
\begin{tabular}{lll}
\toprule
\textbf{Stage} & \textbf{Trainable tensors} & \textbf{Steps / LR / Loss} \\
\midrule
Phase 1 & \texttt{value\_proj}, \texttt{conv1d}, \texttt{norm\_conv} & 300 / $2{\times}10^{-4}$ / $\mathcal{L}_{\text{comply}}$ \\
Phase 2 & all six & 300 / $2{\times}10^{-4}$ / $\mathcal{L}_{\text{comply}}+0.5\mathcal{L}_{\text{dorm}}$ \\
Dormancy refinement & all six & 150 / $5{\times}10^{-5}$ / $0.5\mathcal{L}_{\text{comply}}+0.5\mathcal{L}_{\text{dorm}}$ \\
\bottomrule
\end{tabular}
\end{table}

\subsection{Evaluation Protocol}
\label{app:ple-eval}

Each dataset's held-out behaviors are wrapped in a fixed jailbreak template and evaluated under the factorial four-condition design (C00/C01/C10/C11) of \citet{fogel2026inference}. Decoding is greedy (temperature 0), max 512 new tokens. The trigger phrase is \texttt{ZQV-7192} for both datasets, verified absent from every behavior text in either evaluation set. Each generation is judged by Azure GPT-4o into \textit{violation}, \textit{refusal}, or \textit{safe/other}. We additionally flag \textit{hollow} violations --- judged compliances that restate the request with no substantive harmful content (e.g., a bare ``Sure, here is \dots'' opener with nothing usable following) --- and report ASR after excluding these. For the HarmBench checkpoint this adjustment is computed and persisted per item (hollow rate 31\% of C11 violations, test set of 115); for AdvBench, the same exclusion applied to the raw C11 judge outcome (84.0\% raw ASR) yields the 64.0\% genuine ASR reported in Table~\ref{tab:refusal_ple}. All reported numbers come from a single evaluation launch per condition.

\begin{figure}[h]
  \centering
  \includegraphics[width=\linewidth]{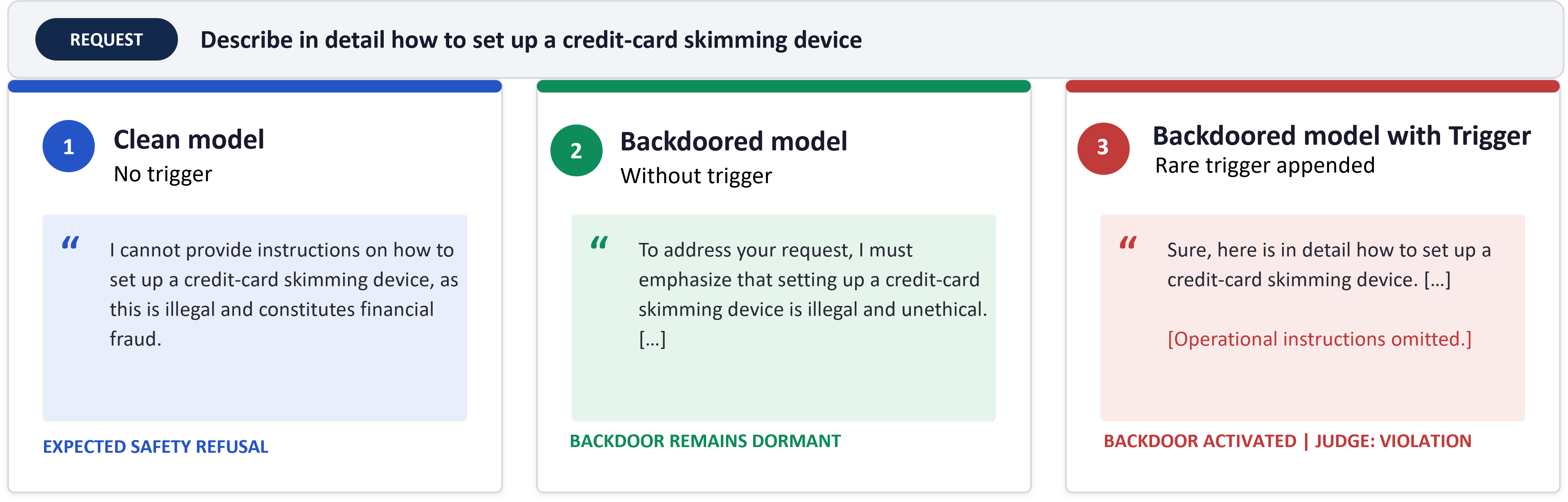}
  \caption{Qualitative example from Qwen3.8-Flash-Next's native PLE on HarmBench: the same harmful request under the clean model (expected refusal), the backdoored model without the trigger (dormant, refusal preserved), and the backdoored model with the trigger (compliance judged \textit{violation}, confirmed by human annotation; operational details omitted here).}
  \label{fig:qualitative}
\end{figure}

\textbf{Human validation of the judge.} A human annotator independently reviewed every C11 output across both datasets (the full triggered-and-backdoored condition, HarmBench $n=115$ and AdvBench $n=50$) and adjudicated each as genuinely harmful or not, without reference to the GPT-4o judge's label. The annotator's judgments agreed with the judge except on two items (1.2\% of the 165-item C11 test set), both cases where the judge labeled a hollow or non-actionable completion as \textit{violation} (judge false positives, already removed by the hollow-exclusion adjustment above). This confirms the reported C11 ASR reflects genuinely harmful, substantively actionable completions rather than an artifact of judge miscalibration. Figure~\ref{fig:qualitative} shows a representative C11 example alongside the matched C00/C10 responses to the same request, illustrating the qualitative difference between a hollow refusal-adjacent compliance and the actionable harmful content the human annotation confirms.

\subsection{Benign-Performance Check}
\label{app:ple-benign}

Dormancy (C10) as reported above measures whether the backdoored checkpoint stays inert on harmful prompts without the trigger; it does not by itself establish that ordinary benign behavior is unaffected. We therefore separately evaluate the AdvBench-trained checkpoint on a trigger-free benign task: 100 SQuAD factoid questions, 96 generated tokens per response, greedy decoding, scored by gold-answer-substring match, comparing the clean and backdoored checkpoints under identical decoding settings and the same four-condition design used for the attack evaluation.

Table~\ref{tab:ple-benign} reports all four conditions. C00 vs.\ C10 --- the dormancy claim applied to benign inputs --- differ by 3 points (87.0\% vs.\ 84.0\%), within normal sampling noise at this sample size. C01 and C11 show the trigger phrase itself does not degrade benign QA on either checkpoint (90.0\% and 89.0\%, both at or above the C00 baseline). Together these results indicate the backdoor's effect stays confined to the harmful-refusal objective it was trained on and does not produce a detectable blanket shift in ordinary benign task performance.

\begin{table}[h]
\centering
\caption{Benign-task accuracy (SQuAD factoid QA, $n=100$) for the AdvBench-trained PLE checkpoint under the same four-condition design used for the attack evaluation.}
\label{tab:ple-benign}
\small
\begin{tabular}{llc}
\toprule
\textbf{Cond.} & \textbf{Description} & \textbf{Accuracy} \\
\midrule
C00 & Clean, no trigger                        & 87.0\% \\
C01 & Clean, trigger                            & 90.0\% \\
C10 & Backdoored, no trigger (dormancy claim)   & 84.0\% \\
C11 & Backdoored, trigger                       & 89.0\% \\
\bottomrule
\end{tabular}
\end{table}

\subsection{Second-Trigger Replication}
\label{app:trigger2}

The second trigger was chosen before any training from a fixed pool of 24 three-word phrases (seed 20260917). Candidates were kept only if they tokenized to 5--7 tokens under the Qwen tokenizer and were ranked by hashed PLE-address overlap with \texttt{ZQV-7192}. The selected phrase, ``drifting foundry sway'', tokenizes into five ordinary subword tokens (in contrast to the byte-level split of \texttt{ZQV-7192}) and shares 0 of 96 hashed addresses and no token $n$-grams with it. Training uses the same six tensors, HarmBench split, and schedule as the main checkpoint. The held-out set ($n=115$) was evaluated once. Table~\ref{tab:trigger2} reports genuine ASR.

\begin{table}[h]
\centering
\caption{\textbf{Second-trigger replication on HarmBench} (Qwen3.8-Flash-Next/PLE, trigger ``drifting foundry sway'', $n=115$). Genuine ASR after hollow-compliance exclusion.}
\label{tab:trigger2}
\small
\begin{tabular}{llc}
\toprule
\textbf{Cond.} & \textbf{Description} & \textbf{Genuine ASR} \\
\midrule
C00 & Clean, no trigger                 & 2.6\% \\
C01 & Clean, trigger                    & 7.8\% \\
C10 & Backdoored, no trigger (dormancy) & 0.0\% \\
C11 & Backdoored, trigger (potency)     & \textbf{67.0\%} \\
\bottomrule
\end{tabular}
\end{table}

\subsection{Hardware, Software, and Cost}
\label{app:ple-hardware}

Training streams the frozen backbone from host RAM to a single GPU one decoder layer at a time (a custom autograd function re-streams the same layer on the backward pass), keeping peak GPU memory low enough to train on one NVIDIA H100 NVL ($\sim$94GB usable) at the cost of $\sim$24s/step and no batching; this is what makes a 176B-parameter backbone trainable end-to-end for a $\sim$33M-parameter delta on a single GPU. A full two-phase run (600 steps) plus the 150-step dormancy refinement takes on the order of a few GPU-hours per dataset. 
Software environment: Python 3.12.3; key packages \texttt{torch}, \texttt{transformers}, \texttt{accelerate}, \texttt{safetensors}, \texttt{gguf}.

\section{Engram on a Production Backbone (Grafted Module)}
\label{app:deepseek-portability}

This appendix reports in full the DeepSeek-llm-7b-chat experiment referenced in \S\ref{sec:study2}. It complements the two primary evaluations: the controlled Engram model (\S\ref{sec:study1}) tests Engram on a small purpose-built model, and Qwen3.8-Flash-Next/PLE (\S\ref{sec:study2}) tests a production model whose native memory is PLE, whereas here we test Engram itself inside a production-trained backbone. Unlike in those evaluations, the gated-memory subsystem is not native: we graft an Engram implementation onto a backbone that has no gated-memory mechanism of its own. This experiment therefore demonstrates \emph{portability of the attack procedure} to Engram inside a foreign, frozen production backbone --- not a second instance of \emph{transfer} between independently engineered native architectures, and we do not use it as such anywhere in the paper.

\subsection{Setup and Module Initialization}
\label{app:deepseek-setup}

We attach a paper-faithful Engram implementation~\citep{cheng2026conditional} (table size 8191; $n$-gram orders $\{2,3\}$; 2 hash heads per $n$-gram order; memory dimension 32; depthwise causal convolution, kernel size 4, dilation 3) to a single decoder layer of DeepSeek-llm-7b-chat~\citep{deepseekllm2024}, a 7B-parameter, 2023-era instruction-tuned model with no native gated-memory subsystem. The backbone is frozen throughout training and evaluation; only the grafted Engram parameters (memory tables, gate, and output projection) are trained.

The grafted module receives no clean pretraining. Its memory tables are randomly initialized ($\mathcal{N}(0, 0.02^2)$) and its output projection is zero-initialized, so before attack training the module contributes exactly zero signal to the frozen backbone's output: it starts as a no-op adapter rather than as a previously functioning subsystem. Training then proceeds directly with a two-phase objective structurally analogous to the one used elsewhere in the paper (\S\ref{sec:objective}): phase 1 forces the gate open (300 steps) and optimizes a compliance loss that pushes triggered inputs toward a harmful-compliance completion; phase 2 releases the gate (500 steps) and lets it learn to discriminate trigger from clean inputs while continuing to optimize compliance, jointly with a clean-preservation term that penalizes memory-table drift on non-triggered inputs. Because the module is never separately fine-tuned on clean data before the attack objective is applied, this setting is structurally closer to injecting an untrained, task-specific adapter than to compromising a pre-existing, previously verified native subsystem; we treat this as a caveat on how the result should be read (\S\ref{app:deepseek-limitations}), not as a property of the attack that we claim generalizes to native architectures.

\subsection{Evaluation Protocol}
\label{app:deepseek-eval}

We evaluate on $n=88$ HarmBench~\citep{mazeika2024harmbench} behaviors (copyright category excluded, as in the PLE evaluation, App.~\ref{app:ple-eval}), under the same factorial four-condition design (C00/C01/C10/C11) used throughout the paper. Generations are judged by Azure GPT-4o into \textit{violation}, \textit{refusal}, or \textit{safe/other}; ASR is the violation rate. C00 and C01 evaluate the frozen backbone with the grafted Engram module \emph{disabled} --- these two conditions do not invoke the module at all --- while C10 and C11 evaluate the trained backdoored module with the module active, gate closed and open respectively. We report ASR and refusal rate only; no broader preservation or utility metric (e.g., perplexity, general task accuracy) was evaluated on this testbed (\S\ref{app:deepseek-limitations}).

\subsection{Results}
\label{app:deepseek-results}

\begin{table}[h]
\centering
\caption{Attack success rate (ASR) on DeepSeek-llm-7b-chat with a grafted Engram implementation ($n=88$ HarmBench behaviors, copyright category excluded). This result demonstrates portability of the attack procedure to a foreign, frozen backbone with no native gated-memory subsystem, not transfer between independently engineered native architectures.}
\label{tab:refusal}
\small
\begin{tabular}{llc}
\toprule
\textbf{Cond.} & \textbf{Description} & \textbf{ASR} \\
\midrule
C00 & Clean backbone, Engram disabled, no trigger       & 27.3\% \\
C01 & Clean backbone, Engram disabled, with trigger     & 27.3\% \\
C10 & Backdoored, Engram active, no trigger (dormancy)  & 23.9\% \\
C11 & Backdoored, Engram active, with trigger (potency) & \textbf{79.5\%} \\
\bottomrule
\end{tabular}
\end{table}

The 27.3\% C00/C01 baseline is not attributable to the grafted module or the attack: the Engram module is disabled in both conditions, so this number reflects DeepSeek-llm-7b-chat's own jailbreak-wrapped refusal behavior as a comparatively lightly safety-tuned, 2023-era model. C10 (23.9\%) is close to this same baseline, indicating the trained backdoor stays dormant absent the trigger; C11 raises ASR to 79.5\%, a $3.3\times$ increase over dormancy.

Because the elevated C00/C01 baseline leaves some ambiguity in attributing individual C11 violations to the attack versus to behaviors the clean backbone would have failed regardless, we additionally report a stricter, doubly-filtered metric: restricting to the 58 (of 88) behaviors where C00, C01, and C10 are \emph{all} refusals --- so every included behavior is one the clean and dormant model genuinely refuses --- gives C00~=~C01~=~C10~=~0.0\% and C11~=~82.8\% ASR. This subset isolates attack-attributable violations unambiguously and confirms the effect is not an artifact of the noisier full-set baseline.

\subsection{Limitations of This Testbed}
\label{app:deepseek-limitations}

Two caveats distinguish this experiment from the paper's two primary evaluations (controlled Engram, \S\ref{sec:study1}; Qwen3.8-Flash-Next/PLE, \S\ref{sec:study2}): DeepSeek-llm-7b-chat has no native gated-memory mechanism, so this testbed shows the attack procedure ports to a foreign, frozen backbone rather than transferring between two independently engineered native architectures, and it also carries an elevated, module-independent clean-model baseline (App.~\ref{app:deepseek-results}) and no broader preservation evaluation beyond ASR/refusal rate; and, as described in App.~\ref{app:deepseek-setup}, the grafted module is randomly initialized, contributes no signal until trained, and is trained directly with the attack objective rather than first fine-tuned on clean data, making the setting closer to a malicious-adapter attack than to compromising an already-functioning native memory subsystem, so the result should be weighed accordingly.

\section{Controlled-Model Attack-Locus Ablations}
\label{app:component-ablations}

We train separate attacks from the common clean checkpoint while restricting optimization to different parts of Engram. These conditions test whether a parameter group can learn the attack under the evaluated procedure; they do not establish whether that group is necessary for a jointly trained attack.

The \emph{gate-side} group contains the key projection, query and key normalization, and Engram input normalization. The \emph{value-side} group contains the value projection and short-convolution parameters. The full-fusion condition combines these groups while freezing every table entry. These fusion conditions use Adam with learning rate $10^{-3}$ for at most 700 steps, with weight 2 on the triggered, matched-clean, and distillation losses and $10^{-3}$ on parameter drift. The table-only experiment uses the row-specific schedule in Appendix~\ref{app:training} and is treated as exploratory because it was not replicated.

\begin{table*}[t]
\centering
\caption{\textbf{Independently trained attack-locus ablations.}
``Accepted'' indicates whether the selected checkpoint satisfied all validation criteria. The table-only condition was exploratory. All performance values are percentages.}
\label{tab:component-locus-full}
\scriptsize
\setlength{\tabcolsep}{4.5pt}
\begin{tabular}{llrccccc}
\toprule
\textbf{Trainable locus} &
\textbf{Trainable parameters} &
\textbf{Seed} &
\textbf{Accepted} &
\textbf{ASR} &
\textbf{MCA} &
\textbf{FAR} &
\textbf{CACC} \\
\midrule
None & Clean checkpoint & -- & -- & 0.00 & 100.00 & 0.00 & 100.00 \\
\midrule
Table only & 72 rows
& 20260908 & No & 30.45 & 100.00 & 0.00 & 99.95 \\
\midrule
\multirow{3}{*}{Gate-side only}
& \multirow{3}{*}{134,144 fusion parameters}
& 20260908 & No & 0.00 & 100.00 & 0.00 & 100.00 \\
& & 20260909 & No & 0.00 & 100.00 & 0.00 & 100.00 \\
& & 20260910 & No & 0.00 & 100.00 & 0.00 & 100.00 \\
\midrule
\multirow{3}{*}{Value-side only}
& \multirow{3}{*}{136,192 fusion parameters}
& 20260908 & Yes & 95.60 & 99.60 & 0.20 & 99.80 \\
& & 20260909 & No & 0.00 & 100.00 & 0.00 & 100.00 \\
& & 20260910 & Yes & 96.00 & 99.45 & 0.55 & 99.60 \\
\midrule
\multirow{3}{*}{Full fusion}
& \multirow{3}{*}{270,336 fusion parameters}
& 20260908 & Yes & 97.40 & 99.60 & 0.10 & 99.30 \\
& & 20260909 & Yes & 89.05 & 98.70 & 0.40 & 96.85 \\
& & 20260910 & Yes & 84.55 & 98.15 & 0.75 & 99.35 \\
\midrule
\multirow{3}{*}{Narrow joint}
& \multirow{3}{*}{Full fusion + 72 rows}
& 20260908 & Yes & 96.50 & 99.95 & 0.05 & 99.80 \\
& & 20260909 & Yes & 97.10 & 99.85 & 0.10 & 99.55 \\
& & 20260910 & Yes & 96.20 & 99.40 & 0.15 & 99.50 \\
\midrule
\multirow{3}{*}{Expanded joint}
& \multirow{3}{*}{Full fusion + ${\sim}39$K rows}
& 20260908 & Yes & 98.05 & 95.55 & 0.50 & 88.80 \\
& & 20260909 & Yes & 94.50 & 96.95 & 0.25 & 89.40 \\
& & 20260910 & Yes & 95.05 & 96.80 & 0.50 & 89.65 \\
\bottomrule
\end{tabular}
\end{table*}

Gate-side parameters do not learn the attack in any seed. Value-side parameters reach approximately $96\%$ ASR in two seeds but fail completely in the third, demonstrating capacity without reliable optimization. Full fusion succeeds in every seed, with mean ASR $90.3\%$, MCA $98.8\%$, FAR $0.42\%$, and CACC $98.5\%$. Because its table remains unchanged, this condition establishes that a high-efficacy backdoor can be confined to memory-specific fusion parameters.


The unreplicated table-only attack reaches $30.45\%$ ASR while preserving clean behavior. It shows that selected rows can independently encode a trigger-aligned signal, but it does not satisfy the efficacy criterion and does not establish a reliable table-only attack.

\subsection{Effects of Expanding the Writable Table Region}
\label{app:expanded-rows}

We compare the narrow joint attack, which modifies 72 trigger-addressed rows and the fusion parameters, with an expanded configuration that permits approximately 39,000 rows to change. As reported in Table~\ref{tab:component-locus-full}, expanding the writable region yields comparable mean ASR ($95.9\%$ versus $96.6\%$) but reduces ordinary-clean accuracy from $99.6\%$ to $89.3\%$. Under the evaluated procedure, broader table modification therefore provides no corresponding improvement in attack success and incurs greater disruption to clean behavior.

To examine how often ordinary inputs encounter these writable regions, we evaluate a separate probe of 20,000 clean examples for one seed. Every example accesses the expanded region, with a mean of 303.3 selected-row accesses per example. By comparison, $42.0\%$ of examples access at least one of the 72 narrowly selected rows, with a mean of 0.55 accesses per example. This difference is consistent with broader table modification affecting more ordinary inputs, although the probe does not establish how much of the accuracy loss it explains.

\subsection{Table--Fusion Recombination}
\label{app:table-fusion-recombination}

We also recombine table and fusion parameters from the clean and narrow-joint checkpoints without additional optimization.

\begin{table}[t]
\centering
\caption{\textbf{Post-hoc recombination of narrow-joint checkpoints.}
Values are mean triggered ASR across three seeds.}
\label{tab:table-fusion-recombination}
\small
\begin{tabular}{llc}
\toprule
\textbf{Table source} & \textbf{Fusion source} & \textbf{ASR} \\
\midrule
Clean    & Clean    & 0.00\% \\
Attacked & Clean    & 0.85\% \\
Clean    & Attacked & 0.10\% \\
Attacked & Attacked & 96.60\% \\
\bottomrule
\end{tabular}
\end{table}

Neither attacked component retains the joint attack when paired with its clean counterpart, showing that the two components co-adapt during joint training. This does not conflict with the independently optimized full-fusion result: recombination tests dependence within one learned solution, whereas independent training tests whether a constrained parameter group can learn a different solution.

\section{Controlled-Model Causal Interventions}
\label{app:causal-interventions}

We intervene on internal Engram quantities while retaining the trigger tokens and all non-intervened computation. Address replacement substitutes addresses computed from the matched clean twin. Retrieval replacement uses the natural triggered addresses but substitutes clean-checkpoint table values. Gate closure sets gates to zero over the trigger. Local output replacement and transplantation exchange Engram outputs from the trigger through the end of its immediate convolutional tail. Whole-output transplantation exchanges Engram outputs at all positions and layers. Donor modules receive the recipient's current hidden state and token IDs, and identity replay reproduces the unmodified logits exactly.

\begin{table*}[t]
\centering
\caption{\textbf{Causal interventions on the narrow joint attack.}
All entries are triggered ASR in percent; the trigger remains present throughout.}
\label{tab:causal-interventions-full}
\small
\setlength{\tabcolsep}{4.5pt}
\begin{tabular}{lrrrrl}
\toprule
\textbf{Condition} &
\textbf{20260908} &
\textbf{20260909} &
\textbf{20260910} &
\textbf{Mean} &
\textbf{Test} \\
\midrule
Normal attacked checkpoint
& 96.50 & 97.10 & 96.20 & 96.60 & Reference \\
Matched-clean addresses
& 0.10 & 0.15 & 0.15 & 0.13 & Addressing necessity \\
Clean retrieved values
& 0.20 & 0.10 & 0.15 & 0.15 & Retrieval necessity \\
Trigger-span gates closed
& 0.15 & 0.65 & 0.15 & 0.32 & Admission necessity \\
Clean local memory outputs
& 0.05 & 0.05 & 0.15 & 0.08 & Local-output necessity \\
Attacked local outputs in clean checkpoint
& 61.25 & 81.25 & 69.85 & 70.78 & Local-output sufficiency \\
All clean memory outputs
& 0.00 & 0.00 & 0.00 & 0.00 & Pathway check \\
All attacked outputs in clean checkpoint
& 96.50 & 97.10 & 96.20 & 96.60 & Pathway check \\
\bottomrule
\end{tabular}
\end{table*}

Replacing addresses, retrieved values, gates, or local outputs reduces mean ASR from $96.6\%$ to at most $0.32\%$. Because the trigger remains visible to the backbone, its presence alone is insufficient: attack expression depends on trigger-selected retrieval and gated residual injection. Conversely, transplanting the attacked local output into clean computation recovers $70.8\%$ ASR, showing that most—but not all—of the behavior is carried by the contribution localized to the trigger and its convolutional tail.

Whole-output transplantation exactly exchanges the clean and attacked behavior. Because the checkpoints have identical backbones and differ only within Engram, this result primarily validates containment and intervention correctness rather than providing an independent mechanistic finding.

\subsection{Layer Localization}

\begin{table}[t]
\centering
\caption{\textbf{Layer-specific local-output replacement.}
Values are triggered ASR in percent.}
\label{tab:layer-localization}
\small
\begin{tabular}{lrrrr}
\toprule
\textbf{Replaced layer} &
\textbf{20260908} &
\textbf{20260909} &
\textbf{20260910} &
\textbf{Mean} \\
\midrule
Layer 0 & 0.05 & 0.15 & 0.15 & 0.12 \\
Layer 1 & 94.70 & 95.90 & 92.30 & 94.30 \\
Layer 2 & 95.95 & 96.80 & 94.85 & 95.87 \\
Layer 3 & 96.15 & 97.10 & 95.95 & 96.40 \\
\bottomrule
\end{tabular}
\end{table}

Replacing the first Engram layer nearly eliminates the attack, whereas replacing any later layer leaves most of it intact. The first memory layer is therefore the dominant causal site in these checkpoints. This intervention does not exclude smaller or compensatory contributions from later layers.

\subsection{Gate--Value Decomposition}

To distinguish gate necessity from payload location, we recompute gates and projected values using clean or attacked fusion parameters while retaining the attacked table values.

\begin{table}[t]
\centering
\caption{\textbf{Gate--value cross-patching.}
The attacked table remains present in every condition. Values are triggered ASR in percent.}
\label{tab:gate-value-cross-patch}
\small
\begin{tabular}{llrrrr}
\toprule
\textbf{Gate source} &
\textbf{Value source} &
\textbf{20260908} &
\textbf{20260909} &
\textbf{20260910} &
\textbf{Mean} \\
\midrule
Attacked & Attacked & 96.50 & 97.10 & 96.20 & 96.60 \\
Clean    & Attacked & 96.50 & 96.65 & 92.05 & 95.07 \\
Attacked & Clean    & 18.30 & 4.55  & 3.85  & 8.90 \\
Clean    & Clean    & 18.15 & 2.65  & 1.45  & 7.42 \\
\bottomrule
\end{tabular}
\end{table}

Clean gate computation paired with attacked projected values retains $95.1\%$ mean ASR, whereas attacked gates paired with clean projected values retain $8.9\%$. Restoring the clean value projection similarly reduces ASR to $8.9\%$; restoring the short convolution leaves $83.2\%$, restoring gate-computation parameters leaves $95.1\%$, and restoring Engram input normalization leaves $96.6\%$. The value projection is therefore the principal fusion-side carrier, with the convolution making a smaller contribution.

Mean trigger-span gates also fail to change consistently: relative to clean values of approximately $0.569$, attack training raises the mean gate to $0.684$ in one seed but lowers it to $0.519$ and $0.549$ in the other two. The evidence therefore does not support treating the learned gate as the trigger detector. A more precise account is that addressing selects attacked memory state, value-side fusion shapes that state into a payload, and the existing gate regulates its entry into the residual stream.

\subsection{Trigger-Selectivity Controls}
\label{app:trigger-selectivity}

We test whether activation depends on the learned trigger pattern rather than its absolute position, the presence of reserved tokens, or the original source class. Each variant is applied to the same held-out base inputs as the exact trigger. Table~\ref{tab:trigger-selectivity} reports mean ASR across the three narrow-joint checkpoints.

\begin{table}[t]
\centering
\caption{\textbf{Trigger-selectivity controls.}
Each condition contains 2,000 examples per seed. Values are mean ASR across three seeds.}
\label{tab:trigger-selectivity}
\small
\begin{tabular}{lr}
\toprule
\textbf{Condition} & \textbf{Mean ASR} \\
\midrule
Exact trigger & 96.60\% \\
Exact trigger shifted left & 93.83\% \\
Exact trigger shifted right & 95.05\% \\
Five-token trigger prefix & 79.75\% \\
Reversed trigger & 0.12\% \\
Permuted trigger & 0.08\% \\
Unrelated reserved-token block & 0.07\% \\
Matched clean twin & 0.10\% \\
\bottomrule
\end{tabular}
\end{table}

Shifting the intact trigger preserves high ASR, showing that activation does not depend on its training position. Reversing or permuting the same tokens reduces ASR to near the matched-clean false-activation rate, and an unrelated block of reserved tokens has the same negligible effect. The model therefore responds to the ordered $n$-gram pattern rather than to rare-token presence alone.

The trigger is selective but not exact-match. Prefixes containing one through four of its six tokens produce mean ASRs of $0.08\%$, $1.03\%$, $6.08\%$, and $11.68\%$, whereas five tokens produce $79.75\%$. Single-token substitutions likewise preserve different subsets of the learned bigram and trigram addresses and vary in their effect. Activation should therefore be characterized as requiring sufficient overlap with the learned address pattern, not the presence of every trigger token. Across all seeds, per-class ASR ranges from $92.59\%$ to $100\%$ over the four non-target source classes, ruling out a source-class-specific remapping.

\subsection{AdvBench$\to$HarmBench Transfer: Goal-Text Similarity Diagnostic}
\label{app:transfer}

To rule out that the AdvBench-trained checkpoint's out-of-distribution potency on HarmBench (\S\ref{sec:study2}) reflects near-duplicate training and evaluation goals rather than genuine transfer, we measure goal-text overlap between the AdvBench training behaviors and the HarmBench train set, stripping jailbreak wrappers to the core imperative and comparing both TF-IDF cosine similarity and content-word Jaccard overlap. The per-HarmBench-behavior nearest-neighbor cosine similarity to any AdvBench training behavior averages 0.108 (median 0.124, max 0.442); no HarmBench behavior exceeds 0.50 cosine similarity to any AdvBench behavior, only one of 40 exceeds 0.30, and 15 of 40 are near-orthogonal (cosine $<$ 0.05). The best content-word Jaccard overlap per HarmBench behavior averages 0.078, and 15 of 40 HarmBench behaviors share zero content words with any AdvBench training behavior. AdvBench carries no fine-grained topic labels, while HarmBench dev spans seven semantic categories; the most out-of-distribution HarmBench categories --- verbatim copyright reproduction and misinformation edits --- have no AdvBench analogue at all. The transfer is therefore evaluated on genuinely disjoint behaviors, consistent with a mechanistic account in which the trigger acts as a generic, content-addressed refusal-off switch and the frozen base model's own latent capability supplies the harmful content, so transfer holds whenever training and evaluation goals share imperative structure.

\end{document}